\documentclass[runningheads]{llncs}

\usepackage{graphicx}
\usepackage[utf8]{inputenc}
\usepackage{color,soul}
\usepackage{comment}
\usepackage{multirow}
\usepackage[normalem]{ulem}
\useunder{\uline}{\ul}{}
\usepackage{listings}
\usepackage{hyperref}
\usepackage{float}
\usepackage{soul}
\usepackage{multicol}

\usepackage{amsmath,amssymb,amstext}

\usepackage[linesnumbered,ruled,vlined]{algorithm2e}

\SetCommentSty{mycommfont}

\SetKwInput{KwInput}{Input}                
\SetKwInput{KwOutput}{Output}              

\usepackage{amssymb, longtable, array}

\newcolumntype{H}{>{\raggedright\arraybackslash}m{\dimexpr(((\linewidth/11)*3)-2\tabcolsep-2\arrayrulewidth)}}
\newcolumntype{T}{>{\centering\arraybackslash}m{\dimexpr((\linewidth/11)-2\tabcolsep-2\arrayrulewidth)}}

\begin{document}

\title{Data Lake Ingestion Management}
%
%

\author{Yan ZHAO\inst{1,2,4} \and Imen MEGDICHE\inst{1,3} \and Franck RAVAT\inst{1,2} }
%
%

\institute{Institut de Recherche en Informatique de Toulouse, IRIT-CNRS (UMR 5505), \and Université Toulouse 1 Capitole, Toulouse, France, \and
Instiut National Unviersitaire J-F Champollion, ISIS Castres, France, \and 
Centre Hospitalier Universitaire (CHU) de Toulouse, France
\email{firstname.lastname@irit.fr}}
\maketitle              
\begin{abstract}
Data Lake (DL) is a Big Data analysis solution which ingests raw data in their native format and allows users to process these data upon usage. Data ingestion is not a simple copy and paste of data, it is a complicated and important phase to ensure that ingested data are findable, accessible, interoperable and reusable at all times. Our solution is threefold. Firstly, we propose a metadata model that includes information about external data sources, data ingestion processes, ingested data, dataset veracity and dataset security. 
Secondly, we present the algorithms that ensure the ingestion phase (data storage and metadata instanciation).
Thirdly, we introduce a developed metadata management system whereby users can easily consult different elements stored in DL. 

\end{abstract}

\section{Introduction}
Data Lake (DL), previously emerged only as a data repository \cite{dixon_pentaho_2010}, is today one of the most popular big data analytics solutions. A DL ensures data ingestion, storage of raw and processed data, preparation of data by different users and consumption of data for different types of analyzes \cite{ravat2019data}.

The major issue of DLs is that they can easily turn into Data Swamps (DS). A DS, contrary to a DL, is invisible and inaccessible to users. It is impossible for data analysts to find, access, retrieve or even reuse needed data effectively \cite{alserafi2016towards}. To meet this lack, metadata management is emphasized for efficient big data analytics \cite{alserafi2016towards,hai2016constance,walker2015personal}, notably for data lineage. 
Data lineage allows analysts to have confidence and trustworthiness in data by providing information about data life-cycle in a DL from raw sources to different usages.

In addition, DLs are characterized by 'schema on read': raw data are ingested without transformation and are processed only when needed. Contrary to traditional ETL (Extract, Transform, Load) processes used for data warehousing and essentially dedicated to the transformation of structured data, ingestion processes of DL face the following specific challenges: connecting to different sources (Databases, Web servers, Emails, IoT, FTP, etc), integrating different data types (structured, semi-structured, unstructured), integrating ‘small’ data transformation (File format change, File compression, Merging multiple small files), supporting multiple ingestion modes (batch, real-Time, one-time load), capturing changes and replicating them in the DL.

To meet these different challenges, it is essential to lean the ingestion processes of a DL associated with an efficient metadata management. As a matter of fact, data ingestion, as the first phase of data life-cycle in a DL, is the foundation of the whole big data analytics process. Current researches only partially meet this problematic.
Some works focus only on one data type (structured \cite{halevy2016goods}, semi-structured \cite{alserafi2016towards}, unstructured \cite{sawadogo2019metadata}), an ingestion mode (batch \cite{alserafi2016towards,sawadogo2019metadata} and real-time \cite{gupta2018data}) or a specific domain \cite{walker2015personal}. Some works address metadata partially \cite{halevy2016goods,alserafi2016towards}. 
Some works introduced different types of metadata without proposing a formalized meta model \cite{quix2016metadata}. 
To the best of our knowledge, there is no work that proposes a metadata model to fully characterize the ingestion process: data sources, ingested data and ingestion processes.

Therefore, in this paper, we propose a data lake metadata management system focusing on data ingestion. Our contributions are (i) a metadata model that includes different categories of metadata generated during the ingestion phase, (ii) metadata feeding mechanisms, and (iii) a tool that allows users to consult metadata to easily interoperate ingested real datasets and reuse them.

The paper is organized as follows, in section \ref{sec:metadatamodel}, we propose a complete metadata model related to the ingestion phase. In section \ref{sec:dataingestion}), we introduce different algorithms to populate the metadata system. In section \ref{sec:implementation}, we present how to find useful information through a web application 
And in section \ref{sec:discussion}, we discuss different solutions of metadata management and data ingestion of DLs.

\section{Metadata Model for Data Ingestion}
\label{sec:metadatamodel}

Data ingestion is the first phase of the data life-cycle in a DL. This phase is the basis of DL management, it cannot be just a copy and paste of raw data for the reason that finding, accessing, interacting and reusing ingested data are all perceived as user needs. In addition, trust and confidence are also expected from users for data exploitation.

Therefore, metadata must be integrated to facilitate future use of DL and they must be collected starting from the ingestion phase. Hence, we present in this section an extension of our metadata model \cite{ravat2019metadata} with precise and complete metadata for the ingesting phase. The extended metadata model is shown in Fig.~\ref{fig:model}. Our model has the following advantages: 

\begin{figure}[t!]
	\includegraphics[width=\textwidth]{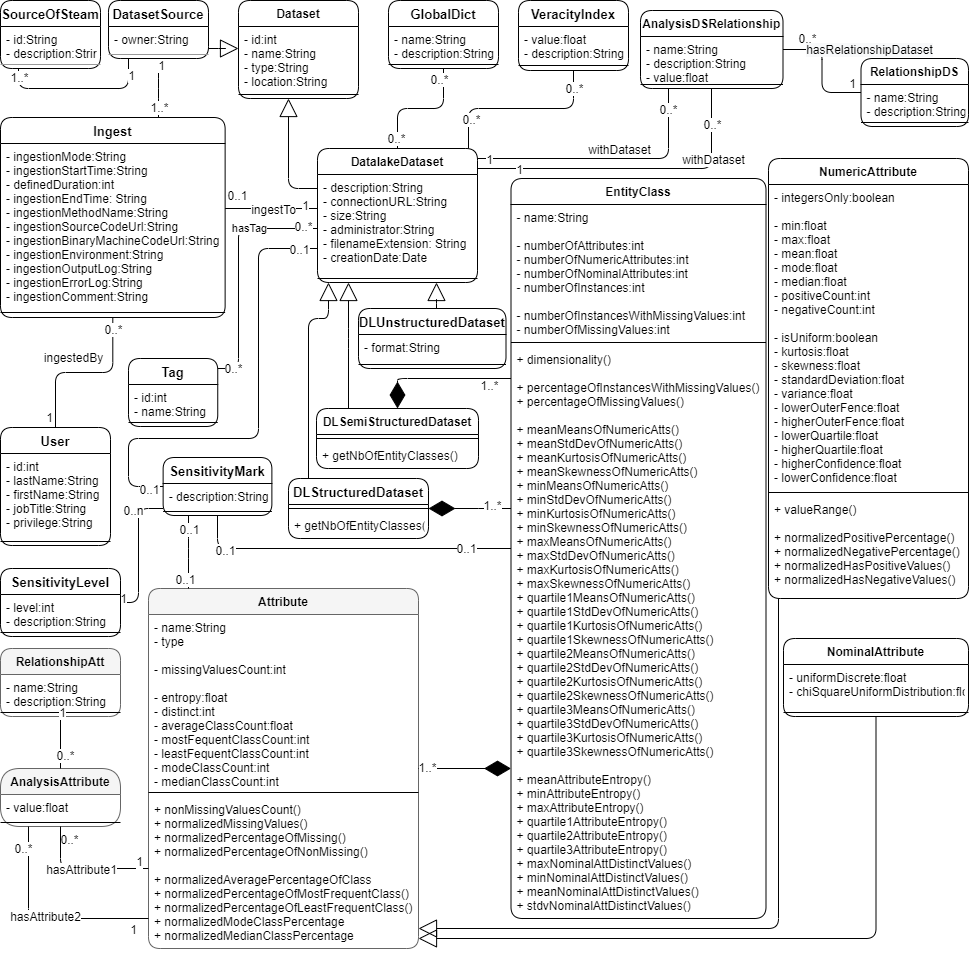}
	\caption{Conceptual Metadata Schema} 
	\label{fig:model}
\end{figure}

\begin{itemize}
    \item It provides \textbf{external data sources information} (class \textit{DatasetSource}) including data source name, type, location, administrator and owner, in addition, source of stream information can be added for IoT (Internet of Things) data. So that users can confirm the origin of the stream data and regroup data from the same source easily.
    \item It provides \textbf{data ingestion process information} (class \textit{Ingest} and its association relationships) including source code, execution details and links to the user, the upstream and downstream datasets. So that users can find the involved datasets and understand how data are ingested, reuse the ingestion process and check or modify the process when errors occurred during the ingestion phase. Note that for batch and real-time data, the start time (\textit{ingestionStartTime}) and end time (\textit{ingestionEndTime}) is the real start and end time of an ingestion process, however, the duration (\textit{definedDuration}) is only required in case of real-time ingestion. 
    \item It provides \textbf{ingested data information} including basic characteristics (class \textit{DatalakeDataset}); \textit{schematic metadata} (classes \textit{DLStructuredDataset, \\ DLSemiStructuredDataset, DLUnstructuredDataset} and their components) that help users to easily understand the structure of data; \textit{semantic metadata} (classes \textit{Tag}, attribute \textit{DatalakeDataset.description}) that help users understand the meaning of data. 
    \item It provides \textbf{dataset veracity information}  (classes \textit{VeracityIndex}) that help users have a first glimpse of data quality. The veracity index is a composite index which combines different measures \cite{rubin2013veracity}: (i) objectivity/subjectivity, (ii) truthfulness/deception, (iii) credibility/implausibility.
    \item It provides \textbf{data security information} (classes \textit{SensitivityMark}, \textit{SensiticityLevel}) to protect sensitive or private data. Users can mark sensitive level for datasets, tables or attributes to limit data access.
    \item It provides  \textbf{relationships between datasets} (classes \textit{RelationshipDS, AnalysisDSRelationship}) so that when users search a dataset, they can find all the relevant data to enrich their analyzes. There are some predefined relationships that can be automatically detected by the system such as the similarity, correlation, containment and logical cluster. In addition, users can define relationships such as a common subject.
    \item It provides \textbf{global metadata} (class \textit{GlobalDict}) which contains DL logs, shared information or public dictionaries among different datasets.
\end{itemize}

The presented meta-model is dedicated to the ingestion phase. It is a complete model which includes information about data sources, all characteristics of the ingested datasets and the ingesting processes. In addition, our model responds to both real time and batch ingestion. The model builds a good base for the data lake management and ensures the good efficacy and high efficiency for different users who work with the DL. 

\section{Data Ingestion Processes}
\label{sec:dataingestion}

The ingestion process in DL consists of two sub-processes: (i) store data in the data lake and (ii) complete and store corresponding metadata. We use algorithms to explain how to ensure the two sub-processes. 

The sub-process \textbf{store data in the data lake} has as an objective to ingest source data into DL by different operation modes and store basic metadata. 
To do so, different steps need to be ensured: 
(i) Establish the source dataset connection and instantiate data source metadata (class \textit{DatasetSource}), the connection information is entered manually. For this step, the function \textit{connectDataSource()} in Algo.~\ref{algo:ingestDataset} is used.
(ii) According to the ingestion mode (real-time/batch), store new dataset and its metadata (class \textit{DatalakeDataset}) as well as ingestion process metadata (class \textit{Ingest}). For this step, the function \textit{ingestDataset()} in Algo.~\ref{algo:ingestDataset} is called.

The sub-process \textbf{complete and store corresponding metadata} has objective of facilitating the future data analysis by helping users to find, access, interoperate the needed datasets by giving users a first vision of dataset semantic and schematic information of chosen datasets. To achieve this objective, we need to complete and store not only the information of each single dataset but also the relationships among different datasets (in no particular order and can be parallel):

\begin{flushleft}
\scalebox{0.7}{
\begin{minipage}{1.42\linewidth}
\begin{algorithm}[H]
\DontPrintSemicolon
    \tcc{establish source dataset connection}
    \SetKwFunction{connectDataSource}{connectDataSource}
    \SetKwProg{Fn}{Function}{:}{}
    \Fn{\connectDataSource{$dataSourceConnection$, $dataSourceType$, $dataSourceLocation$, $dataSourceName$, $dataSourceOwner$, $sourceOfSteam$, $owner$}}{
        \If{tryConnectDataSource($dataSourceConnection$, $dataSourceLocation$)}{
            $datasetSource$ $\leftarrow$ instantiateClass('DatasetSource', createProperties('name', $dataSourceName$, 'type', $dataSourceType$, 'location', $dataSourceLocation$, 'owner', $owner$)) \;
            \If{$sourceOfSteam$}{
                createRelation('DatasetSource-SourceOfSteam', $datasetSource$, $sourceOfSteam$) \;
            }
            return $datasetSource$
        }\Else{
            return None
        }
    }
    \tcc{ingest dataset}
    \SetKwFunction{ingestDataset}{ingestDataset}
    \SetKwProg{Fn}{Function}{:}{}
    \Fn{\ingestDataset{$datasetSource$, $ingestionComment$}}{
        $ingestionProcess$, $ingestedDataset$ $\leftarrow$ ingestDatasetInDL($datasetSource$) \;
        $ingest$ $\leftarrow$ instantiateClass('Ingest', getPropertiesIngest($ingestionProcess$), 'comment', $ingestionComment$) \;
        createRelation('DatasetSource-Ingest', $datasetSource$, $ingest$) \;
        $datalakeDataset$ $\leftarrow$ instantiateClass('DatalakeDataset', getPropertiesDatalakeDataset($ingestedDataset$)) \;
    }
\caption{Functions used to ingest datasets into a data lake}
\label{algo:ingestDataset}
\end{algorithm}
\end{minipage}
}
\end{flushleft}

\begin{itemize}
    \item \textbf{Instantiating definitional metadata} concerns semantic and schematic metadata. Semantic metadata include a list of tags (class \textit{Tag}) and a description (\textit{DatalakeDataset.description}) of datasets and can be manually input by users. Schematic metadata of structured and semi-structured databases consist of the database model with tables/entities and attributes information. For unstructured datasets, the data format is detected. This task can be implemented according to the functions \textit{instantiatingSemanticMetadata()} and \textit{instantiatingSchematicMetadata()} which are introduced in Algo.~\ref{algo:definitionalmatadata}.
    \item \textbf{Instantiating security metadata} consists of the security levels of different data granularity (dataset, table, column, etc). The security metadata is not necessary, they can be manually input or not, the default value is zero which means that all users can access the dataset. However, for some special domains, for example, hospitals, to protect the personal data of patients, the sensitivity metadata are mandatory. Concerning the population of metadata database, security metadata is input by users to ensure the protection of sensitive data.
    \item \textbf{Instantiating inter-metadata} concerns various relationships among datasets which are important for data finding, reuse and interoperability in a data lake. The inter-metadata can be automatically generated by the system (for instance, similarity and coherence between one dimension tables) or input manually by users. Note that the generation of inter-metadata is not limited in this sub-process, for example, the provenance information can be instantiated during the data storage phase as we mentioned previously. To populate the inter-metadata, the functions \textit{calculateRelationships()} and \textit{inputRelationships()} can be called (see Algo.~\ref{algo:intermatadata}).
    \item \textbf{Instantiating veracity metadata} is about calculating the veracity composite index which includes objectivity, truthfulness and credibility \cite{rubin2013veracity} of datasets.   
\end{itemize}

\begin{flushleft}
\scalebox{0.7}{
\begin{minipage}{1.42\linewidth}
\begin{algorithm}[H]
\DontPrintSemicolon
    \tcc{instantiating definitional metadata - semantic}
    \SetKwFunction{instantiatingSemanticMetadata}{instantiatingSemanticMetadata}
    \SetKwProg{Fn}{Function}{:}{}
    \Fn{\instantiatingSemanticMetadata{$datalakeDataset$, $datalakeDatasetDesc$, $datalakeDatasetTags$}}{
        addAttributes('DatalakeDataset', $datalakeDataset$, createProperties('description', $datalakeDatasetDesc$)nh) \;
        \tcc{store tags for datalakeDataset}
        $allTags[]$ $\leftarrow$ getAllTags() \;
        \ForEach{$t$ $\subset$ $datalakeDatasetTags$}{
            \If{$t$ in $allTags[]$}{
                $tag$ $\leftarrow$ getTag($t$)
            }\Else{
                $tag$ $\leftarrow$ instantiateClass('Tag', createProperties('name', $t$))\;
            }
            createRelation('DatalakeDataset-Tag', $DatalakeDataset$, $tag$)
        }
    }
    \tcc{instantiating definitional metadata - schematic}
    \SetKwFunction{instantiatingSchematicMetadata}{instantiatingSchematicMetadata}
    \SetKwProg{Fn}{Function}{:}{}
    \Fn{\instantiatingSchematicMetadata{$datalakeDataset$}}{
        \If{($datalakeDataset.type$ = "structured" or "semi-structured")}  
        {
            $entities[] \leftarrow$ getEntityClasses($datalakeDataset$)\;
            \ForEach{$e \subset entities$}{
                $atts[] \leftarrow$ getAttributes($e$)\;
                $entityClassProperties \leftarrow$ createProperties('name', getEntityName($e$), getEntityStatistics($atts[]$)) \; \tcp{function getEntityStatistics() returns an array including the name and value of each statistical metadata}
                $entityClass \leftarrow$ instantiateClass('EntityClass', $entityCLassProperties$) \;
                createRelation('DatalakeDateset-EntityClass', $datalakeDataset, entityClass$) \;
                \ForEach{$att \subset atts[]$}{
                    \If{getAttType(att) = 'numeric'}{
                        $numericAttributeProperties \leftarrow$ createProperties('name', getAttName($att$), getNumericAttStat($att$))
                        $attribute \leftarrow$ instantiateClass('NumericAttribute', $numericAttributeProperties$)\;
                    }
                    \Else{
                        $nominalAttributeProperties \leftarrow$ createProperties('name', getAttName($att$), getNominalAttStat($att$)) \;
                        $attribute \leftarrow$ instantiateClass('NominalAttribute', $nominalAttributeProperties$)\;
                    }
                    createRelation('EntityClass-Attribute', $entityClass, attribute$) \;
                }
                \tcc{For each predefined RelationshipAtt we calculate the value of relationship between attributes}
                $analysisAttributes[] \leftarrow$ getAnalysisAttribute(atts[], $relationshipAtts[]$) \; 
                \ForEach{$an \subset analysisAttributes[]$}{
                    $relationArr \leftarrow$ instantiateClass('AnalysisAttribute', createProperties('value', $an.value$) \;
                    createRelation('AnalysisAttribute-Attribute', $relationArr, an.attribute1$) \;
                    createRelation('AnalysisAttribute-Attribute', $relationArr, an.attribute2$) \;
                    createRelation('AnalysisAttribute-RelationshipAtt', $relationArr, an.relationshipAtt$) \;
                }
                
            }
        }
        \Else{
        	addAttributes('DatalakeDataset', $datalakeDataset$, getDatasetFormat($datalakeDataset$))
        }
    }
\caption{Functions used to complete definitional metadata}
\label{algo:definitionalmatadata}
\end{algorithm}
\end{minipage}
}
\end{flushleft}

\begin{flushleft}
\scalebox{0.7}{
\begin{minipage}{1.42\linewidth}
\begin{algorithm}[H]
\DontPrintSemicolon
    \tcc{for chosen dataset, calculate automatically relationships}
    \SetKwFunction{calculateRelationships}{calculateRelationships}
    \SetKwProg{Fn}{Function}{:}{}
    \Fn{\calculateRelationships{$datalakeDataset$}}{
        \tcc{For each predefined RelationshipDS we calculate the value of relationship between datasets}
        $analysisDSRelationships[] \leftarrow$ getAnalysisDSRelation($datalakeDataset, datalakeDatasets[], relationshipDSs[]$) \; 
        \ForEach{$anDs \subset analysisDSRelationships[]$}{
            $relationDs \leftarrow$ instantiateClass('AnalysisDSRelationship', createProperties('value', $anDs.value$)) \;
            createRelation('AnalysisDSRelationship-DatalakeDataset', $relationDs, anDs.datalakeDataset1$) \;
            createRelation('AnalysisDSRelationship-DatalakeDataset', $relationDs, anDs.datalakeDataset2$) \;
            createRelation('AnalysisDSRelationship-RelationshipDS', $relationDs, anDs.relationshipDS$) \;
        }
    }
    \tcc{for chosen datasets, users input a relationship manually}
    \SetKwFunction{inputRelationships}{inputRelationships}
    \SetKwProg{Fn}{Function}{:}{}
    \Fn{\inputRelationships{$datalakeDataset1$, $datalakeDataset2$, $relationshiDS$, $dsRelationshipName$, $dsRelationshipdesc$, $dsRelationshipValue$}}{

        $analysisDSRelationship$ $\leftarrow$ instantiateClass('AnalysisDSRelationship', ) \;
        createRelation('AnalysisDSRelationship-DatalakeDataset', $relationDs, anDs.datalakeDataset1$) \;
        createRelation('AnalysisDSRelationship-DatalakeDataset', $relationDs, anDs.datalakeDataset2$) \;
        createRelation('AnalysisDSRelationship-RelationshipDS', $relationDs, anDs.relationshipDS$) \;
    }
\caption{Functions used to complete inter metadata}
\label{algo:intermatadata}
\end{algorithm}
\end{minipage}
}
\end{flushleft}

\section{A Metadata Management System}
\label{sec:implementation}


To facilitate data discovery in a data lake, we have implemented a metadata management system\footnote{\url{https://github.com/yanzhao-irit/data-lake-metadata-management-system}}. This system contains a user-oriented web application and a database to store metadata. The application,  developed mainly in JavaScript and HTML/CSS, has an ergonomic graphical interface that allows users to easily interact with metadata. The database is a graph database (Neo4j) which has the advantages of good scalability, flexibility and maturity.

To illustrate the usage of the metadata application and set up an available example, we ingested different types of real open-source datasets\footnote{\url{https://github.com/yanzhao-irit/data-lake-metadata-management-system/tree/main/example-metadata}} (see table \ref{tab:datasets}). The metadata graph database contains 2315 nodes and 2775 relationships. Where, regarding data ingestion, there are 7 nodes of \textit{DatasetSource}, 7 nodes of \textit{Ingest}, 55 nodes of \textit{EntityClass}, 1197 nodes of \textit{Attribute} and 74 nodes of \textit{Tag} etc. Regarding the relationship detection, we created 3 datasets based on the \textit{CHSI cancer} dataset and calculated the relationships between the these datasets (1 node of \textit{RelationshipDS} and 3 nodes of \textit{AnalysisDSRelationship}) and the relationships between attributes in these datasets (5 nodes of \textit{RelationshipAtt} and 62 nodes of \textit{AnalysisAttribute}).

\begin{table}[]
\centering
\resizebox{\textwidth}{!}{%
\begin{tabular}{llllll}
\hline
\textbf{Dataset}            & \textbf{Description}                    & \textbf{Type} & \textbf{Entities} & \textbf{Columns} & \textbf{Size} \\ \hline
MIMIC                   & a freely accessible critical care database & relational database & 40   & 534 & 6.2 GB    \\
Chest X-Ray Images      & Chest X-ray images (anterior-posterior)    & images              &  &     & 1.15 GB   \\
CHSI cancer             & health indicators for the US counties      & csv file            & 11   & 587 & 15 MB     \\
Country Vaccinations    & COVID-19 World Vaccination Progress        & csv file            & 1    & 15  & 2.92 MB   \\
Breast Cancer Wisconsin & diagnostic of breast cancer wisconsin      & csv file            & 1    & 32  & 122.27 KB \\
Fetal health classification & classification of the health of a fetus & csv file      & 1                 & 22               & 22.35 KB      \\
Lung Cancer             & lung cancer dataset with four indicators   & csv file            & 1    & 7   & 1.66 KB  
\end{tabular}%
}
\caption{Ingested datasets}
\label{tab:datasets}
\end{table}

\begin{figure}[t!]
    \centering
	\includegraphics[width=\textwidth]{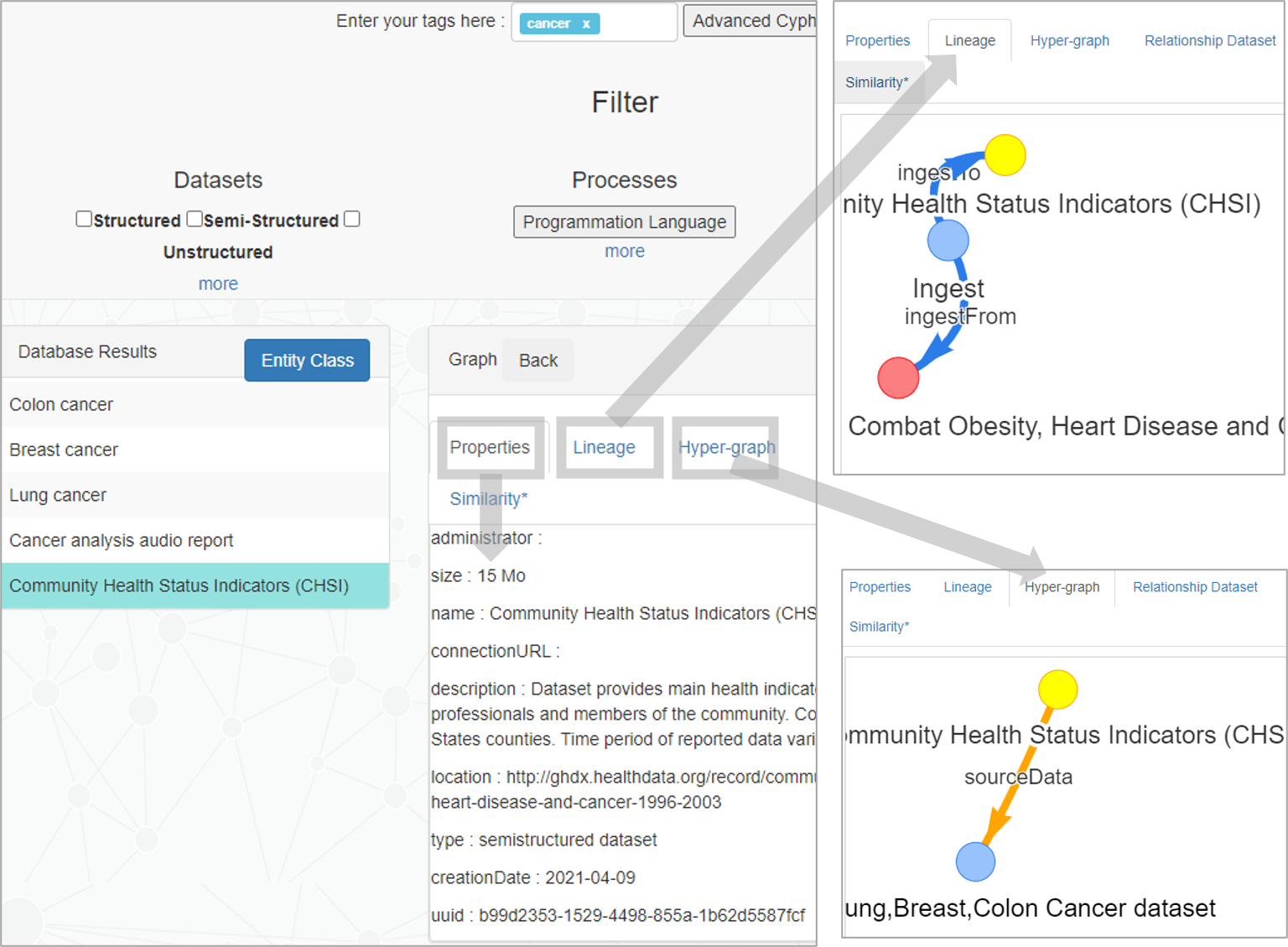}
	\caption{Dataset information and ingestion process information} 
	\label{fig:usecase1}
\end{figure}

\begin{figure}[t!]
    \centering
	\includegraphics[width=\textwidth]{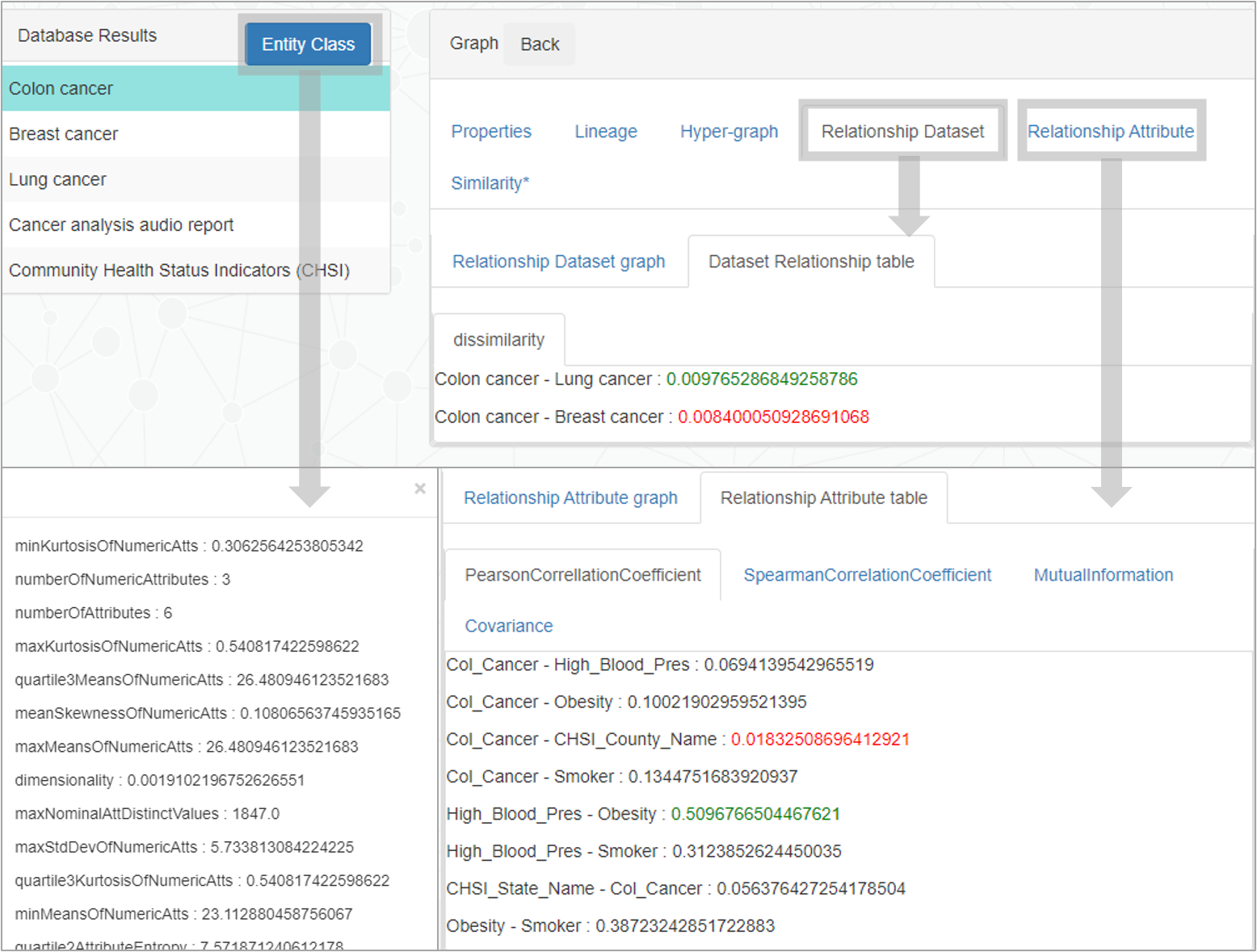}
	\caption{Dataset basic information and ingestion process information} 
	\label{fig:usecase2}
\end{figure}

Moreover, the system allows users to easily find, consult datasets that are ingested in the DL via a web application. Data analysts who work with the DL can use the application to find useful information. To validate the interest of this metadata, we asked end users to use our tool. For instance, a data analyst needs to start a project concerning cancer analysis. He wants to find in the DL if there are existing datasets about cancer, he can easily search the key word \textit{'cancer'} in the application. Then he can see a list of datasets whose name, description or tags contain the key word (see the left result in Fig.~\ref{fig:usecase1}). Moreover, he can chose different datasets then click on available tabs to get more details of chosen datasets to see if they are useful for his project or not. For instance, he can find the properties and lineage information of dataset to see what is the dataset about and how and when it is ingested in the DL (see Fig.~\ref{fig:usecase1}). In addition, he can find more details of datasets attributes information and the relationships between different datasets (see Fig.\ref{fig:usecase2}).

\section{Discussion}
\label{sec:discussion}

The data ingestion in DL is not a simple copy of data. The process is much more complicated as illustrated throughout the paper. Bringing a complete metadata system is the best solution to prepare a DL for data analytics, data lineage and making it compliant with FAIR principles \cite{noauthor_fair_nodate}. To the best of our knowledge, there is no complete solution which includes all metadata of ingested datasets, data ingestion processes and developed metadata management tool.

\begin{figure}[!h]
    \centering
	\includegraphics[width=\textwidth]{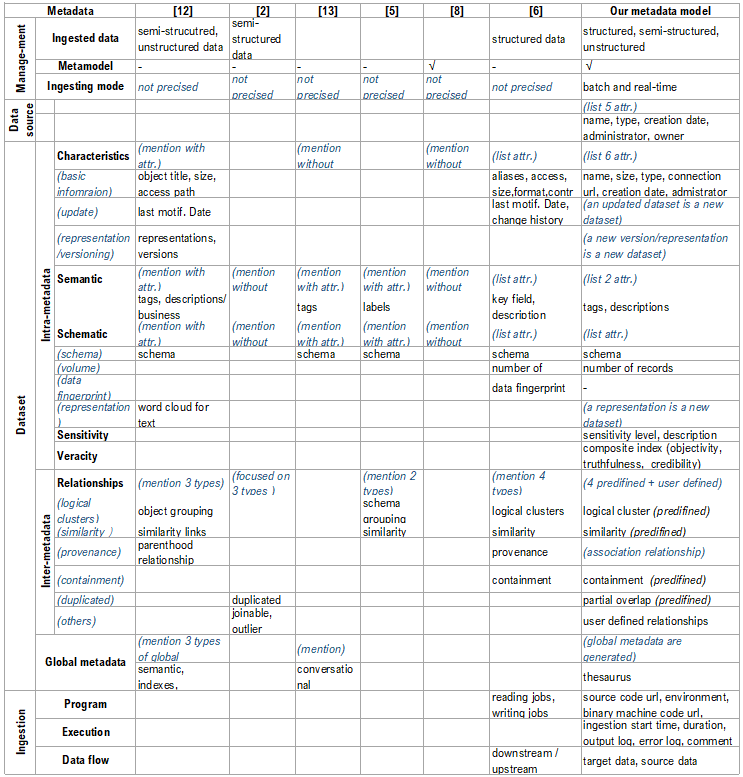}
	\caption{A comparative table of Metadata for the ingestion phase.} 
	\label{fig:tableComparaison}
\end{figure}

Regarding the metadata management and in order to discuss our proposal compared to the state of the art approaches, we summarize in Fig.~\ref{fig:tableComparaison} the different types of metadata that should be addressed. This comparison includes the most relevant approaches which studied the metadata for DLs: MEDAL \cite{sawadogo2019metadata} that proposes a not formalized metadata model focusing on semi-structured and unstructured data;
CM4DL \cite{alserafi2016towards} that focuses on the inter-metadata of semi-structured data;
DL-Wrangling \cite{terrizzano2015data} that introduces a not formalized metadata model which contains global metadata;
Constance \cite{hai2016constance} that focuses on data ingestion without proposing a formalized metadata concerning this phase,
GEMMS \cite{quix2016metadata} that introduces a metadata model without including inter-metadata and 
GOODS \cite{halevy2016goods,halevy2016managing} that concerns a post-hoc metadata management.

In Fig. \ref{fig:tableComparaison}, concerning the management aspects, we can observe that our approach is the only one that covers different types of ingested data (structured, semi-structured and unstructured) supported with a metadata model. Regardless its importance, the state of the art approaches do not precise the ingesting mode. Our approach satisfy both batch and real-time data ingestion. Both types of ingestion mode are crucial in data lake environment. 

In addition, it is always important to be able to trace the data source used for the ingestion. External data source metadata allow users to check the provenance of ingested datasets so that they can find the data source in case of updating datasets. Five attributes were dedicated in our model for this subject but we have noticed that the other approaches neglected this point. 

Moreover, concerning datasets metadata, we distinguish three types of metadata: intra-, inter- and global-metadata. Regarding intra-metadata, we have the originality of including sensitivity and veracity metadata to ensure the access control and data governance. Regarding inter-metadata, our model is more complete than GOODS and MEDAL with particularly the possibility for users to define their own dataset relationships. Regarding global metadata, we use a global dictionary for shared information among datasets and it can be simply used by research engines.

What's more, concerning the ingestion process metadata, to the best of our knowledge, there is only GOODS \cite{halevy2016goods,halevy2016managing} which address this point without a formalized processing model. 
For the ingestion process, program, execution and data flow metadata should be saved. The approach of Goods \cite{halevy2016goods} saves reading and writing jobs. In our approach, we are more generic as we save source code URL, binary machine code URL and the whole environment.  GOODS does not track the execution metadata while we back up ingestion start time, duration, output log, error log and further comments. Our approach is then compliant with lineage on processes. 

Regarding the ingestion process, to the best of our knowledge, there are only a few solutions. \cite{alserafi2016towards} introduced metadata management for data ingestion by three phases (ingest, digest and exploit) for batch data. They focused on the extraction of inter-metadata from semi-structured data. \cite{gupta2018data} presented a data ingestion framework which consists of a data collector and a data integrator. They explained different approaches to bring data into a Hadoop data lake without integrating a formalized ingestion processing and a metadata model for this phase. \cite{terrizzano2015data} introduced data ingestion by three steps without a formalized processing.

These data ingestion solutions did not introduce a detailed ingestion processing model. Moreover, they did not include both the aspects of data ingestion processing and the metadata of data ingestion. Therefore, we proposed a more complete metadata model and a data ingestion processing that includes the interaction of the DL system and users. Currently, we are formalizing the other data lake processes (data processing and data access) using algorithms. After having demonstrated the feasibility of our approach, we intend to focus on performance optimizations of integrating a great volume of various types of datasets. 

\section{Conclusion}

Today, data lakes are increasingly promoted as a big data analysis solution. In this paper, we focus on improving the efficacy and the efficiency of data ingestion. To the best of our knowledge, our work is the only one which proposes both a complete metadata model and the underlying processes for the ingesting phase. 

Regarding metadata for data ingestion, we propose a complete model including information on data source, datasets and ingesting process. In addition, for datasets, our model contains inter and intra metadata to facilitate future data analysis. Regarding ingestion processes, we formalize the data ingestion processes including metadata generation using algorithms. Moreover, our ingestion processes are general enough for different structural types of data and for different ingestion modes. We validate our solution by implementing a metadata management system with real datasets and illustrate the system with a use case to show its ease of use and ergonomic interface.

Currently, we are completing our proposal with additional data lake processes (data preparation and data analysis). To be compliant with data lineage, we formalize these processes and identify the associated metadata. After having demonstrated the feasibility of our approach, we intend to focus on performance optimizations of our metadata management system.

\bibliographystyle{splncs04}
\bibliography{bib}

\end{document}